\documentclass[preprint2]{aastex}

\slugcomment{Submitted to ApJ, updated 12/3/04}

\shorttitle{Be Stars in Open Clusters}
\shortauthors{McSwain \& Gies}

\include{comdef}

\begin{document}

\title{A Photometric Method to Search for Be Stars in Open Clusters}


\author{M. Virginia McSwain\altaffilmark{1,2,3} and Douglas R. Gies} 
\affil{Department of Physics and Astronomy, Georgia State University, 
P.O. Box 4106, Atlanta, GA 30302-4106}
\email{mcswain@astro.yale.edu, gies@chara.gsu.edu}

\altaffiltext{1}{Current address: Department of Astronomy, Yale 
University, P.O. Box 208101, New Haven, CT 06520-8101}
\altaffiltext{2}{Visiting Astronomer, Cerro Tololo Inter-American 
Observatory.  CTIO is operated by AURA, Inc.\ under contract to the 
National Science Foundation.}
\altaffiltext{3}{NSF Astronomy and Astrophysics Postdoctoral Fellow}

%%%%%%%%%%%%%%%%%%%%%%%%%%%%%%%%%%%%%%%%%%%%%%%%%%%%%%%%%%%%%%%%%%%%%

\begin{abstract}
We describe a technique to identify Be stars in open clusters using 
Str\"omgren $b$, $y$, and narrow-band H$\alpha$ photometry.  We first 
identify the B-type stars of the cluster using a theoretical isochrone 
fit to the $(b-y, y)$ color-magnitude diagram.  The strongest Be stars are 
easily identified in a $(b-y, y-\rm H\alpha)$ color-color diagram, but 
those with weaker H$\alpha$ emission (classified as possible Be star 
detections) may be confused with evolved or foreground stars.  Here we 
present such photometry plus H$\alpha$ spectroscopy of members of the 
cluster NGC 3766 to demonstrate the accuracy of our technique.  
Statistical results on the relative numbers of Be and B-type stars in 
additional clusters will be presented in a future paper. 
\end{abstract}


%%%%%%%%%%%%%%%%%%%%%%%%%%%%%%%%%%%%%%%%%%%%%%%%%%%%%%%%%%%%%%%%%%%%%

% Section 1
\section{Introduction}

Be stars are broadly defined as non-supergiant (luminosity class
III$-$V) B-type stars that show or have shown Balmer emission
\citep{porter2003}, although a few supergiant Be stars are also known
\citep{negueruela2004}.  Their emission is due to a circumstellar disk
that is often highly variable on both short and long time scales.
The emission lines change intensity, and the relative intensity of
their double-peaked profiles changes, over short time scales of days
and even hours.  In addition, the disks themselves have been observed
to disappear and reappear over years or decades \citep{slettebak1988}.
Be stars are well known to be a class of rapidly rotating stars 
\citep*{slettebak1949, slettebak1966, slettebak1992}, and there are three 
possible reasons for their rapid rotation.  They may have been born as 
rapid rotators \citep{zorec1997}, spun up by binary mass transfer 
\citep{pols1991}, or spun up during the main-sequence (MS) evolution of B 
stars \citep{meynet2000, fabregat2000}.

Many observational surveys have searched for Be stars in open clusters 
using multi-band photometry.  \citet{shobbrook1985, shobbrook1987} used 
plots of $c_1$ vs. $\beta$, $c_1$ vs. $V$ magnitude, and $\beta$ vs. $V$ 
to identify Be stars in NGC 3766.  
%Here, $c_1 = (u-v) - (v-b)$, a common Str\"{o}mgren index used to measure 
%the Balmer jump, and $\beta$ is the magnitude from a narrow-band H$\beta$ 
%filter that measures H$\beta$ emission from Be stars.  
Many recent surveys have relied on color-color diagrams, rather than 
Shobbrook's color-magnitude diagrams, to detect Be stars more efficiently.
\citet*{grebel1992} used a technique somewhat similar to ours (\S3) to 
identify Be stars in NGC 330.  They used Str\"omgren $b, y$, and a 
narrow-band H$\alpha$ filter for their observations, and they plotted a 
color-color diagram for the cluster using $b-y$ vs H$\alpha-y$.  They were 
able to verify almost all of their detections with low resolution spectra, 
demonstrating the accuracy of this technique.  Several other surveys 
(\citealt*{grebel1997, dieball1998, keller1999, keller2000, grebel2000, 
pigulski2001}) have used variations of the color-color diagram technique 
to identify Be stars in open clusters.  \citet{grebel1997} emphasizes that 
the number of Be stars detected in this type of surveys actually 
represents a lower limit on the true number of Be stars for several 
reasons.  Most importantly, since the Be state is not permanent, only 
those stars with current disk outbursts will be detectable as Be stars.  
In addition, the number of late type Be stars is likely to be incomplete 
since they are fainter, and they also tend to have weaker H$\alpha$ 
emission than early Be stars.

In this paper, we present an improved photometric technique based on 
$(b-y,y)$ color-magnitude and $(b-y, y-\rm H\alpha)$ color-color diagrams 
to detect Be stars in open clusters.  While our method is very similar to 
that of \citet{grebel1992}, we have enhanced it by using cluster isochrone 
fitting to identify the B-type stars in each cluster.  We have also used a 
fit of the color-color curve to identify H$\alpha$ emitters among the B 
star population.  To demonstrate the accuracy of our technique, we present 
spectra of 20 stars in the cluster NGC 3766 and correlate their H$\alpha$ 
equivalent widths with their $y-\rm H\alpha$ colors.  Our photometry of 
additional clusters will be presented in a future paper, and at that time 
we will use our statistics to discuss the implications on the possible Be 
star formation scenarios.

%%%%%%%%%%%%%%%%%%%%%%%%%%%%%%%%%%%%%%%%%%%%%%%%%%%%%%%%%%%%%%%%%%%%%

% Section 2
\section{Observations}

%% In a manner similar to \objectname authors can provide links to dataset
%% hosted at participating data centers via the \dataset{} command.  The
%% second curly bracket argument is printed in the text while the first
%% parentheses argument serves as the valid data set identifier.  Large
%% lists of data set are best provided in a table (see Table 3 for an example).
%% Valid data set identifiers should be obtained from the data center that
%% is currently hosting the data.

We made photometric observations of the cluster NGC 3766 on 2002 
April 1 using the CTIO 0.9 m telescope with the SITe 2048 CCD.  The 
images were binned using a CCD summing factor of $2\times2$ pixels due 
to the slow readout time of the chip.  Without binning, the chip has a 
plate scale of 0.401\arcsec/pixel, but with binning the plate scale 
increased by a factor of 2.  

We observed the cluster using Str\"{o}mgren $b$ and $y$ filters 
as well as a narrow band H$\alpha$ filter, and we obtained two 
observations of 5 s duration in each filter.  We also observed five 
standard stars (HD 79039, HD 80484, HD 104664, HD 105498, and HD 
128726) from the list of \citet{cousins1987}.  Each standard star was 
observed in each band at a minimum of three different airmasses.  
All of the images were processed in IRAF\footnote{IRAF is
distributed by the National Optical Astronomy Observatory, which is
operated by the Association of Universities for Research in Astronomy,
Inc., under cooperative agreement with the National Science 
Foundation.} using nightly bias and dome flat frames, and we used the 
technique described by \citet{mas92} to compute the instrumental 
magnitudes of each object.  We determined the magnitudes of the 
standard stars using a large aperture of 8\arcsec, and we used a 
smaller aperture for the more crowded cluster stars and later 
performed an aperture correction to transform the target measurements to 
the aperture system of the standards. 

We calibrated our Str\"omgren $b$ and $y$ photometry of NGC 3766 using 
the photometric transformation equations from \citet{henden1982}, but 
the H$\alpha$ magnitudes were somewhat harder to correct since the 
filter system is not universally standard.  We corrected the H$\alpha$
instrumental magnitudes for atmospheric extinction by plotting each
standard star's instrumental magnitude vs.\ airmass, and the slope 
provides a good measure of the principal extinction coefficient.  
Since the color coefficient is difficult to determine without knowing 
the absolute H$\alpha$ magnitudes, and since color is relatively 
unimportant for the response of such a narrow-band filter, we assumed it 
to be negligible within errors.  We also assumed that the second-order 
term is zero.  We determined the constant offset between the instrumental 
H$\alpha$ magnitudes and the true magnitudes by fitting a theoretical 
color-color curve to the MS B stars in the observed
color-color diagram (described in \S 3).  Because the theoretical 
curve assumes that Vega, an A0~V star, is assigned a magnitude of zero, 
the final H$\alpha$ magnitudes are calibrated using this convention.  The 
errors in the magnitudes are from the combined estimated instrumental 
errors, the standard deviation of the aperture correction, and the 
errors in the transformation coefficients.  They are usually $< 0.04$ mag 
in $b$ and $y$, and $< 0.07$ mag in H$\alpha$.  Our photometry of NGC 
3766 is listed in Table 1, which is available online (see Note to Table 
1).

\placetable{table1}

From our images, we also performed astrometry of the cluster to determine 
accurate positions of each star.  Ten medium-bright stars from each 
cluster were selected to be astrometry reference stars, and we matched 
their average $(x,y)$ positions with their right ascension, $\alpha$, and 
declination, $\delta$, from the 2MASS All-Sky Catalog of Point Sources 
\citep{cutri2003}.  We used the IDL routine \textit{astromit.pro}, written 
by R.\ Cornett \& W.\ Landsman, to compute the astrometric solution from 
the reference stars.  With this known solution, the program 
\textit{uit\_xy2ad.pro}, written by R.\ Cornett, B.\ Boothman and J.\ D.\ 
Offenberg, used the list of $(x,y)$ positions of all stars in NGC 3766 and 
output their $\alpha$ and $\delta$.  Using this technique, we obtained 
typical errors of $0\farcs10$ in both $\alpha \cos \delta$ and $\delta$.

In addition, we obtained spectra of 21 members of NGC 3766 with the CTIO 
1.5 m telescope in 2003 March.  These first-order spectra were made with 
the Cassegrain spectrograph, the \#47 grating (831 grooves~mm$^{-1}$, 
blazed at 8000 \AA), a GG495 order sorting filter, and a Loral 
$1200\times800$ CCD.  This arrangement recorded the spectrum over the 
range 5490 -- 6790 \AA ~with a resolving power of $\lambda / \Delta 
\lambda = 1700$.  Exposure times varied between 300 -- 1800 s, and 
resulted in a S/N = 300 pixels$^{-1}$ near H$\alpha$.  We did not remove 
the weak, narrow telluric lines from these spectra so that our 
measurements of the line flux more closely correspond to the photometric 
measurements.  These spectra were reduced to a rectified intensity versus 
heliocentric wavelength format using standard procedures in IRAF.

%%%%%%%%%%%%%%%%%%%%%%%%%%%%%%%%%%%%%%%%%%%%%%%%%%%%%%%%%%%%%%%%%%%%%

% Section 3
\section{Detection Technique}

Figure \ref{colormag} shows a color-magnitude diagram of NGC 3766 
with the theoretical isochrone, assuming solar metallicity, from 
\citet{lejeune2001}.  For this cluster, we used the values of the 
reddening, $E(b-y)$ = 0.15, and unreddened distance modulus, $5\log d-5$ = 
11.4, provided 
by \citet{shobbrook1985} and $\log$ age (years) = 7.16 from the WEBDA 
database for Galactic open clusters\footnote{The WEBDA database is 
maintained by J.-C. Mermilliod and is available online at 
obswww.unige.ch/webda/navigation.html.}.  However, 
\citeauthor{lejeune2001} do not include an ischrone for this cluster age 
in their database, so we used the closest match ($\log$ age = 7.15) in 
Figure \ref{colormag}.  Because their models do not provide 
Str\"omgren magnitudes, we transformed the Johnson $V$ magnitude to 
Str\"omgren $y$ using the transformation
\begin{equation}
y = V - 0.038(B-V)
\end{equation}
\citep{cousins1985}.  To transform $B-V$ to $b-y$, we used
\begin{equation}
B-V = 1.584(b-y) + 0.681m_1 - 0.116
\end{equation}
from \citet{turner1990}.  Not having obtained the $v$ magnitudes for the 
cluster, we used the ZAMS relation from \citet{perry1987} to find 
appropriate values of $m_1$ for different values of $b-y$, and we 
interpolated between them for all other values of $b-y$.

\placefigure{colormag}		% Figure 1

Each point along the theoretical isochrone corresponds to an effective 
temperature, and we defined all objects with temperatures between
$10000-30000$ K as B stars.  Some B-type giants and supergiants may 
be included in this range as a result.  The horizontal dashed line in 
Figure \ref{colormag} represents the maximum magnitude for B-type
stars in the cluster, $y_{B max}$, while the minimum B star magnitude, 
$y_{B min}$, is out of the plot range.  The massive O- and B-type stars 
all lie to the left of the vertical dashed line that represents $E(b-y)$.  
It is important to note that the gap between the faintest stars 
and the horizontal line $y_{B max}$ may indicate a systematic bias in this 
search for Be stars; in NGC 3766, our photometry does not detect the 
faintest late-type B or Be stars.

Before inspecting the observed color-color curve of NGC 3766, we 
created a theoretical color-color curve using colors derived from 
\citet{kurucz1979} model spectra.  These model flux distributions
include a realistic treatment of the H$\alpha$ line profile, and they
have a fine enough resolution to permit a reasonable estimate of the
flux that passes through the narrow-band H$\alpha$ filter (about 5
model wavelength points span the full range of the filter).   We used
the IRAF routine $calcphot$ in the $HST stsdas/synphot$ package to
calculate the $y - {\rm H}\alpha$ and $b - y$ indices for the
Kurucz model fluxes assuming solar metallicity, $\log g = 4.0$ and
4.5, and a temperature range of 7000 -- 30000~K.  We omitted higher
temperature, O-type stars because non-LTE effects become important in
these stars, and Balmer line absorption is stronger than predicted by
the Kurucz LTE models.  We had to adjust the Str\"{o}mgren $b$ and $y$ 
magnitudes from $calcphot$ for the true magnitudes of the A0 V star 
Vega since they differ slightly from zero in the Str\"{o}mgren system 
\citep{gray1998, hauck1998}.  We then interpolated between the two gravity 
models $\log g = 4.0$ and 4.5 for the gravity associated with MS stars in 
the \citet{lejeune2001} models.  The solid line in Figure \ref{kurucz} 
shows the resulting theoretical
color-color curve from the Kurucz model spectra.  Along this curve,
the $y - {\rm H}\alpha$ index rises with $b - y$ color in redder
(cooler) stars, and the curve has a local minimum near $b - y = 0.0$
that corresponds to the maximum H Balmer absorption line strength
found in early A-type stars.  The dashed line in Figure \ref{kurucz}
represents the same Kurucz model colors at a reddening of $E(B-V)$ =
0.5 using the reddening law from \citet{fitzpatrick1999} for an assumed 
value of $R \equiv A(V)/E(B-V) = 3.1$.  Note that for this value of $R$, 
$E(B-V)$ can be converted to the Str\"{o}mgren system using the expression
\begin{equation}
E(b-y) = 0.745 \times E(B-V)
\end{equation}
from \citet{fitzpatrick1999}.

\placefigure{kurucz}	% Figure 2 - Theoretical color-color plot

Figure \ref{kurucz} also includes the color indices derived from
dereddened spectra of 161 stars from the atlas of \citet*{jacoby1984}.
%Diamonds mark MS stars while more luminous stars are indicated by plus
%signs.  
There appears to be good agreement between the reference
color-color relation from the Kurucz fluxes and the observed colors of
MS stars in the B- and A-type spectral range ($b - y =
-0.13$ to 0.25).  The mean difference is $(y - {\rm H}\alpha)_{\rm
obs} - (y - {\rm H}\alpha)_{\rm mod} = 0.01 \pm 0.05$, where the error
indicates the standard deviation.  The most discrepant point is 0.10
mag away from the predicted trend.  This indicates that the reference
curve can be applied to our photometry with relatively small
systematic errors.  However, it does point out the need to be
cautious in identifying Be stars.  Hot, post-MS stars have slightly
larger $y - {\rm H}\alpha$ indices than MS stars since they have
narrower H$\alpha$ absorption profiles due to less pressure broadening
in their atmospheres.  In addition, unreddened MS stars may appear
slightly above the reddened theoretical curve and could be confused
with weak Be stars.  A parabolic fit of the colors of the entire
Jacoby et al.\ sample is shown as a dotted curve, and unreddened
foreground stars will probably be found close to this line in the
observed color-color diagrams.

It was difficult to match the theoretical color-color curve to the 
observed data, however, since the model assumes $y - \rm H\alpha = 0$ 
for A0 V stars (specifically, Vega), but the H$\alpha$ photometry was 
not initially calibrated to such a standard system.  Therefore, we used 
the B star data to find the average difference between our data and
the model, and we offset the H$\alpha$ magnitudes by adding this
difference.  Because the fit was sometimes skewed by Be stars with
significantly larger $y - \rm H\alpha$ values, we removed any stars
more than 2$\sigma$ above the fit and improved the offset.  Here, $\sigma$ 
is the standard deviation of the $y - \rm H\alpha$ residuals from the 
Kurucz relation for the stars in NGC 3766.

The color-color diagram of NGC 3766 is plotted in Figure \ref{colorcolor}, 
which also includes the theoretical color-color curve (solid line) 
transformed according to the adopted reddening $E(b-y)$ for the cluster 
using the reddening law from \citet{fitzpatrick1999}.  The long dashed 
line in the diagram is the parabolic fit for unreddened MS and evolved 
stars from the Jacoby et al.\ sample.  Finally, the vertical dotted 
line in Figure \ref{colorcolor} indicates $E(b-y)$.  

\placefigure{colorcolor}	% Figure 3 - color-color diagram

With theoretical models correlated to both the color-magnitude and 
color-color diagrams, we used the agreement to identify both B stars 
and Be stars among the cluster members.  The B stars lie in the region 
$y_{B min} < y < y_{B max}$ and $b-y < E(b-y)$.  A subset of these B 
stars may be active Be stars, which are distinguished by their line 
emission, often in H$\alpha$, so they will usually appear brighter in 
this band than normal B stars.  Those B stars with $y - \rm H\alpha$ 
more than $2\sigma$ above the theoretical color-color curve (and above 
the unreddened MS fit, to eliminate unreddened foreground stars and 
post-MS stars) are tentatively classified as Be stars.  Each Be star 
candidate is represented with a large diamond in Figures \ref{colormag} 
and \ref{colorcolor}, whereas all other stars are shown as small diamonds.  
Definite Be star detections are those candidates which have a 
significantly larger ($>5\sigma$) $y - \rm H\alpha$ color than the 
cluster's B-type population.  The possible detections are those whose 
colors lie just slightly above the unreddened MS fit.  These possibly 
include some contaminating foreground stars and supergiants, but we
show in \S4 that this population does include some weakly emitting 
Be stars.

%%%%%%%%%%%%%%%%%%%%%%%%%%%%%%%%%%%%%%%%%%%%%%%%%%%%%%%%%%%%%%%%%%%%%

% Section 4
\section{Spectroscopic Test of the Method}

To test the accuracy of this photometric method of identifying Be stars in 
NGC 3766, we performed a spectroscopic investigation of this cluster.  
From a literature search, we found 22 previously identified Be stars in 
NGC 3766.  \citet{mermilliod1982} lists 13 Be stars identified by 
H$\alpha$ line emission in their spectra, and \citet{shobbrook1985, 
shobbrook1987} classified 10 of Mermilliod's stars plus 9 additional stars 
as Be stars based on his own Str\"{o}mgren $uvby$ and H$\beta$ photometry.

Our own color-color diagram of NGC 3766, shown in Figure \ref{colorcolor}, 
revealed only 5 Be stars with definite H$\alpha$ emission (WEBDA \#15, 
264, 240, 88, and 53).  In addition to these, one of the previously 
identified Be stars (\#232) was saturated in our images, and one was not 
in the field-of-view (\#301).  To resolve the discrepancy between the 
published literature and our results, we obtained spectra of these 
previously identified Be stars, excluding only the star outside the 
field-of-view in our CCD images. 

From the spectra of NGC 3766, we found that only 9 of the 21 stars 
exhibited H$\alpha$ emission in our spectra.  One of these emission stars 
(\#232) was indeed the known Be star that is saturated in the photometry.  
We show the relative H$\alpha$ profiles of the other 8 emission stars in 
NGC 3766, offset vertically for clarity, in Figure \ref{haprofiles}.  Five 
of these 8 stars, those with the strongest H$\alpha$ emission, are easily
identified as Be stars in the color-color diagram in Figure 
\ref{colorcolor}.  Therefore, our technique for identifying definite Be 
star candidates with a $b-y$ vs $y-$H$\alpha$ diagram accurately accounts 
for about 63\% of the Be stars in this cluster.  

\placefigure{haprofiles}	% Figure 4 - Ha emission profiles

The three Be stars that did not stand out in the color-color diagram
each had somewhat different H$\alpha$ line profiles from the other Be
stars.  One star (\#1) had a ``shell'' spectrum, i.e., weak emission
in the wings of the line accompanied by stronger central
absorption.  This line profile is characteristic of Be stars observed
nearly edge-on, so that the thick hydrogen disk absorbs an even larger
fraction of the star's flux at H$\alpha$.  The other two Be stars 
(\#27 and \#151 southern component) had only weak H$\alpha$ emission, 
probably from smaller disks, making it more difficult to distinguish 
them from non-emission stars.

After reviewing the results from the spectra, we looked more carefully
at Figure \ref{colorcolor} and noticed a group of stars
occupying a region just slightly above the bulk of the OB stars, but
beneath the unreddened line that we use as the cutoff for Be star
detections.  These may be weak H$\alpha$ emitters, or they may
be foreground stars.  Within the errors of the photometry, this group
is nearly indistinguishable from the non-emitters.  The shell star \#1
was in this group, so it was not detected as a Be star candidate.
However, the weak emitter \#27 was found just above the unreddened
line, so it was identified as a possible Be star by our technique.
The final weak emitter, \#151, was grouped with the non-emitting OB
stars in the color-color diagram.  On the other hand, two stars that
did not show H$\alpha$ emission in the spectra, \#26 and \#204, 
appeared as possible Be star detections above the unreddened line.  
Given that the weakly-emitting Be stars and the non-emitting B star 
regions overlap within the errors of the photometry, it is easy to 
attribute the confusion to the photometry errors.  However, it is worth 
noting that \#26 and \#204 have been classified as Be stars in the past, 
and there was a one year gap between obtaining photometry and spectroscopy 
of this cluster.  Therefore it is also possible to attribute the 
discrepancy between the photometry and spectroscopy to intermittent weak 
emission from these stars.  Therefore we count each of the stars in 
the weakly-emitting region of the color-color diagram as possible Be star 
detections, but we do not include \#1 or \#151 as a detection based on 
the criteria discussed above.

Twelve of the previously identified Be stars in NGC 3766 did not have
active H$\alpha$ emission at the time of our spectroscopic observations.  
We show their H$\alpha$ absorption profiles in Figure \ref{haprofiles2} on 
a slightly different scale than Figure \ref{haprofiles}.  Several of these 
stars (especially \#291, 204, 195, 146, 36, and 4) have absorption line 
profiles that may have a small emission component in the absorption 
line cores due to weak nebular H$\alpha$ emission.  Alternatively, the 
emission-like profile could be an indication of line blending in 
double-line spectroscopic binaries.  This superposition of two absorption 
line profiles can mimic Be star emission profiles; \citet{cote1993} found 
that several stars in their Be star sample had been misclassified as 
emission stars due to their binary nature.  More spectra of these stars 
are needed to confirm if they are indeed spectroscopic binaries that have 
been misclassified as Be stars in the past.

\placefigure{haprofiles2}	% Figure 5 - Ha profiles of non Be stars

To demonstrate the overall accuracy of this technique to identify Be 
stars, we show in Figure \ref{haeqwidth} a plot of the equivalent width, 
$W_\lambda$, of H$\alpha$ versus the $y-\rm H\alpha$ color of the 20 stars 
with both photometry and spectroscopy.  $W_\lambda$ was measured by 
performing a numerical integration from $6529 - 6597$ \AA, the range 
covered by the narrow band H$\alpha$ filter (FWHM = 68 \AA, centered at 
6567 \AA).  The measurement errors are dominated by the placement of the 
continuum and amount to $\delta W_\lambda = \pm 0.22$~\AA ~or better.  
Absorption features have positive equivalent width while emission features 
have negative equivalent width, but the equivalent width may be positive 
for weak emission line stars in which the emission only partially fills 
the photospheric absorption line of H$\alpha$.  We fit the observed 
$W_\lambda$ and $y-\rm H\alpha$ color to the logarithmic relationship
\begin{equation}
y-{\rm H}\alpha = C + 2.5\,\log \left (\frac{{\rm FWHM}_{{\rm 
H}\alpha}
- W_\lambda}{{\rm FWHM}_{{\rm H}\alpha}} \right)
\end{equation}
with the fitting factor $C = 0.264$, and this curve is shown as a solid 
line in Figure \ref{haeqwidth}.  There is significant scatter in the 
observed relationship compared to the theoretical curve, and the boundary
between Be and non-Be stars is unclear, probably due to the variability of 
H$\alpha$ emission over the year between our observations.  However, Be 
stars appear to have $y - {\rm H}\alpha \gtrsim 0.2$ in our calibrated 
system.

\placefigure{haeqwidth}		% Figure 6 - Ha eq widths

%%%%%%%%%%%%%%%%%%%%%%%%%%%%%%%%%%%%%%%%%%%%%%%%%%%%%%%%%%%%%%%%%%%%%

\section{Conclusions}

Our photometric technique for identifying Be stars is very efficient for
finding those stars with strong H$\alpha$ emission, as we have shown for
NGC 3766.  While we correctly identify 63\% of the previously known,
currently active Be stars in NGC 3766, we do not confirm the active Be
status of 7 stars claimed as Be stars in the past.  As our results show,
Be stars with weaker emission are more difficult to identify conclusively
with the color-color diagram method.  Even if the errors in the photometry
were negligible, these stars may be confused with unreddened foreground
stars or supergiants in a color-color diagram.  In some cases it may be
possible to eliminate such foreground stars from the sample based on their
positions relative to the MS in a color-magnitude diagram; alternatively, 
proper motions or radial velocities may be used to separate non-members 
from the cluster sample.  We do not think that B-type supergiants are a 
significant source of contamination since supergiants within the cluster 
will be rejected by the $y_{Bmin}$ constraint in the color-magnitude 
diagram, and the chances of having a background B supergiant in the sample 
are very small unless two distant clusters fall along the same line of 
sight (this is the case with some of the clusters in our forthcoming 
paper).  However, B-type supergiants can be removed from the sample by 
including IR photometry in the observational program since Be star disks 
will create an IR excess that is correlated to their luminosity in 
H$\alpha$ \citep{stee2001}.  If such measurements are not available, 
H$\alpha$ spectroscopy may be required to conclusively identify
weak H$\alpha$ emitters.  

\begin{figure}
\includegraphics[angle=90,scale=0.3]{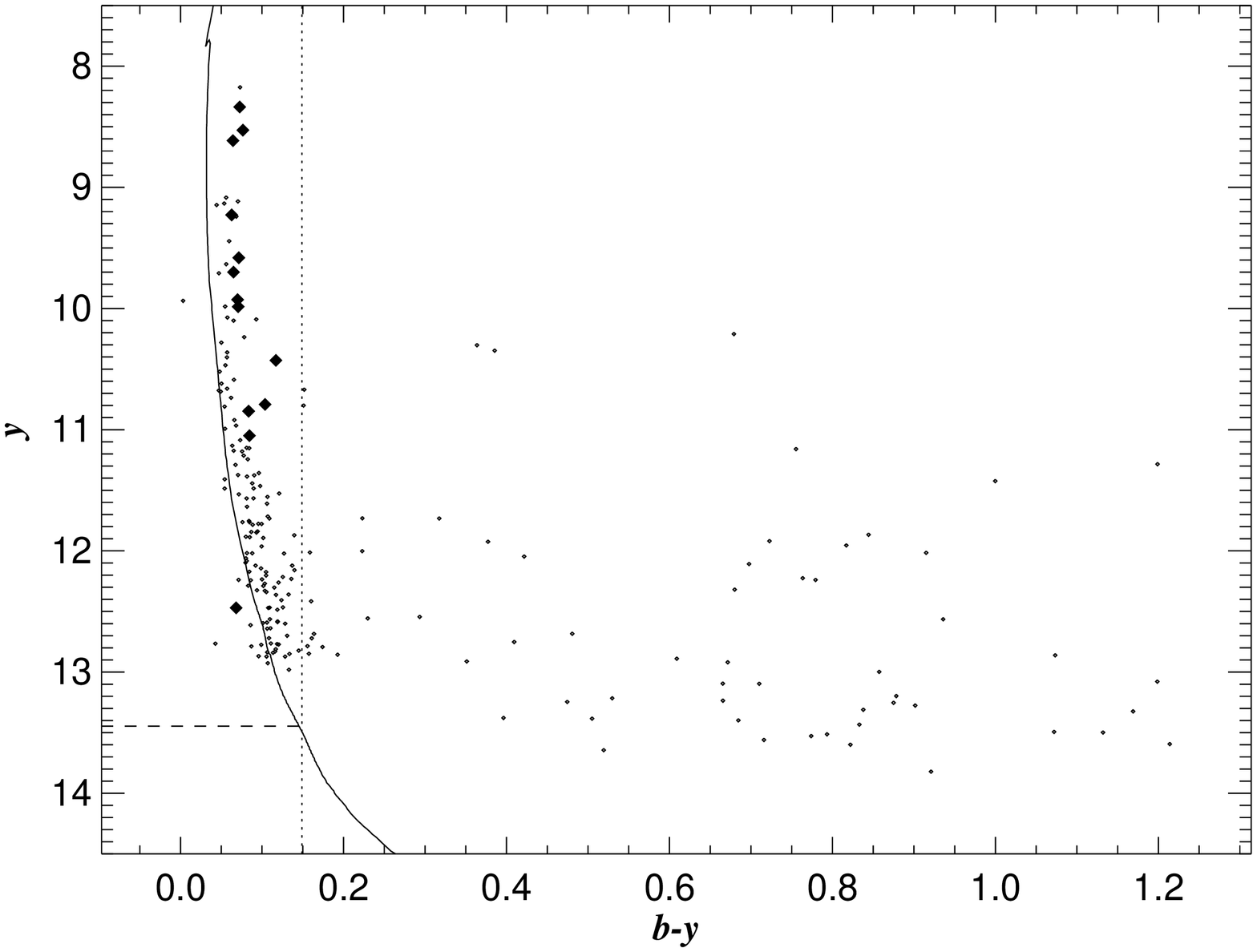}
\caption{
Color-magnitude diagram of the cluster NGC 3766.  The isochrone fit 
(\textit{solid line}), $E(b-y)$ (\textit{dotted line}), and $y_{Bmax}$ 
(\textit{dashed line}) are discussed in the text.  Be stars 
(\textit{large}) are distinguished from all other stars (\textit{small}) 
in the diagram.
\label{colormag}
}
\end{figure}

\begin{figure}
\includegraphics[angle=90,scale=0.3]{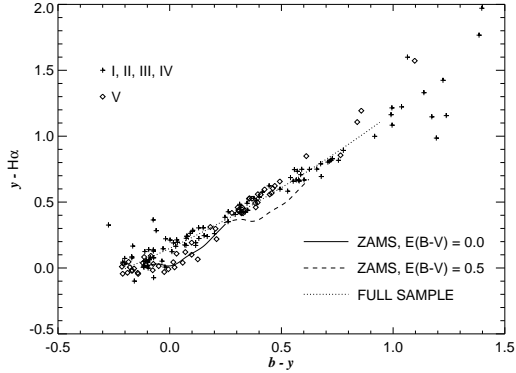}
\caption{
The theoretical unreddened color-color curve (\textit{solid line}) and
reddened curve (\textit{dashed line}) derived from \citet{kurucz1979}
model spectra are plotted.  In addition, the colors of 161 MS
(\textit{diamonds}) and more luminous stars (\textit{plus signs}) from 
the atlas of \citet{jacoby1984} are shown.  The parabolic fit of these 
stars' colors is also shown (\textit{dotted line}).
\label{kurucz}
}
\end{figure}

\begin{figure}
\includegraphics[angle=90,scale=0.3]{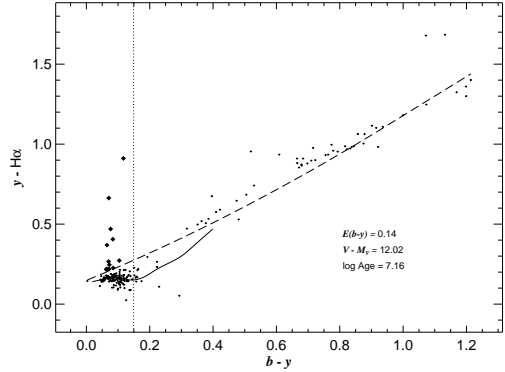}
\caption{
Color-color diagram of the cluster NGC 3766.  The color-color fit from the 
Kurucz model fluxes (\textit{solid line}), the parabolic fit for 
unreddened MS and evolved stars from \citet{jacoby1984} (\textit{dashed 
line}), and $E(b-y)$ (\textit{dotted line}) are also indicated.  The Be 
stars are distinguished in the same format as Figure \ref{colormag}.
\label{colorcolor}
}
\end{figure}

\begin{figure}
\includegraphics[scale=0.5]{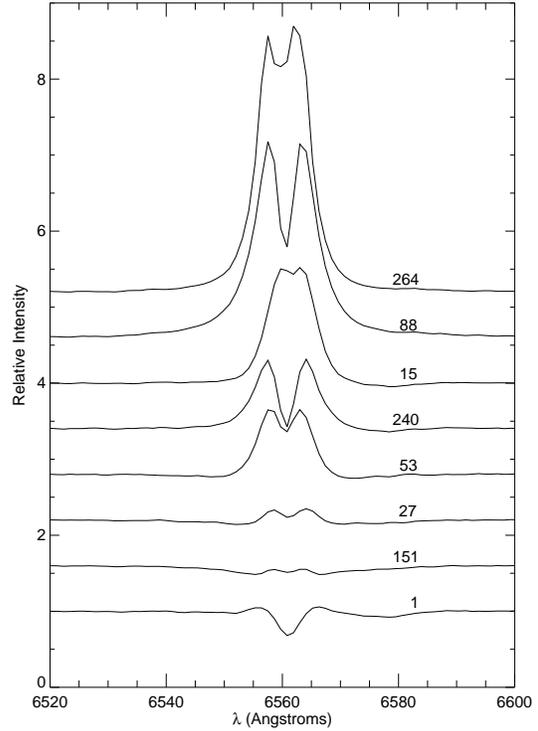}
%\plotone{f4.ps}
\caption{
The relative H$\alpha$ profiles are shown for 8 Be stars in NGC 3766, 
labeled by their WEBDA number.  The relative intensities are offset 
vertically for clarity.
\label{haprofiles}
}
\end{figure}

\begin{figure}
\includegraphics[scale=0.5]{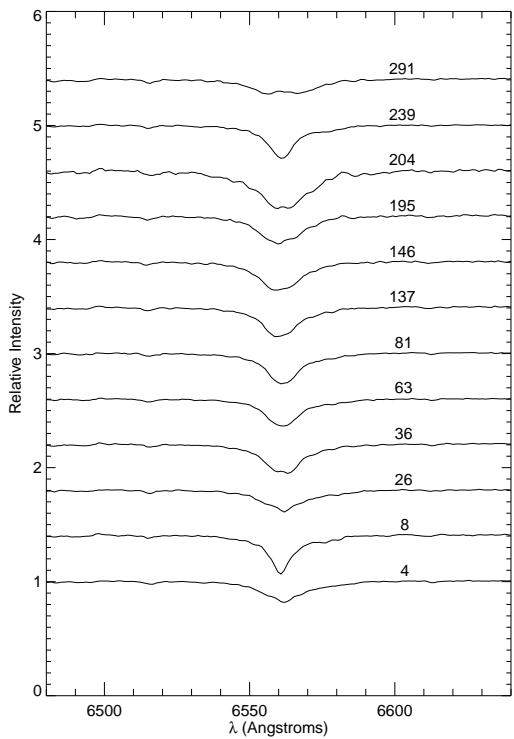}
%\plotone{f5.ps}
\caption{
The relative H$\alpha$ profiles are shown for 12 inactive Be stars in
NGC 3766 in the same format as Figure \ref{haprofiles}.
\label{haprofiles2}
}
\end{figure}

\begin{figure}
\includegraphics[angle=90,scale=0.3]{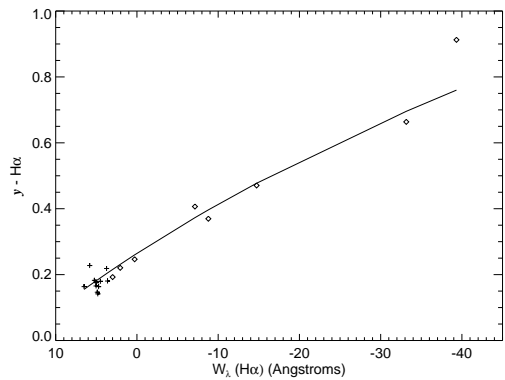}
\caption{
The H$\alpha$ equivalent widths from our spectra are shown plotted
against the observed $y-\rm H\alpha$ color from our photometry for 20
stars in NGC 3766.  Stars with H$\alpha$ emission in their spectra
indicating that they are Be stars (\textit{diamonds}) generally have a 
higher $y-\rm H\alpha$ color index and more negative $W_\lambda$ than 
non-Be stars (\textit{crosses}).  The theoretical logarithmic 
relationship between $W_\lambda$ and $y-\rm H\alpha$ is also shown as 
a solid line. 
\label{haeqwidth}
}
\end{figure}

As Figure \ref{colormag} shows, our photometry of NGC 3766 does not cover 
the entire range of B star magnitudes.  Unless the observational program 
is designed to reach A-type stars, the total number of B, and perhaps Be, 
stars will be underestimated in any search for Be stars.  This is 
especially a concern when measuring the fraction of Be stars relative to 
normal B-type stars because the number of B-type stars increases 
dramatically with later spectral types.  \citet{keller2001} illustrate 
this problem with their deep photometric survey of the clusters $h$ and 
$\chi$ Persei.  They find that more than 30\% of the brightest B stars are 
Be stars, but the Be star fraction decreases to 2\% or less for B5--B9 
spectral types.  Because many other studies of Be stars (e.g.  
\citealt{mermilliod1982, shobbrook1985, grebel1997, fabregat2000}) rely 
upon magnitude limited samples that do not accurately reflect the true Be 
fractions, this source of incompleteness is worth discussion.  However, in
this situation a more accurate Be/B star fraction can be determined by
assuming a typical IMF distribution for the cluster and extrapolating to
estimate the number of late stars omitted.  As \citet{keller2001} show,
the contribution of late-type Be stars is small and may be neglected in
such an extrapolation.

Both \citet{abt1987} and \citet{grebel1997} have discussed another problem 
with short-term photometric studies of Be stars, namely that they only 
detect those Be stars with active emission at the time of the 
observations, and thus only a lower limit of the true Be population is 
found.  This effect can be significant in observational surveys; 
\citet{cote1993} found that 14\% of the known Be stars in their sample of 
B stars from the Bright Star Catalogue did not show active emission.  From 
our own spectra of NGC 3766, we found that only 9 of 21 Be stars discussed 
in the literature had active disks in 2003 March, although we believe that 
at least some of these Be stars have been misidentified in the past.  From 
these two studies, the true fractions of Be stars in a sample could be 
anywhere between $1.2-2.3$ times the number detected.  Only a long term, 
multi-year investigation will approach the complete population of Be stars 
in a given sample.

While our $(b-y, y-\rm H\alpha)$ color-color technique is limited in its 
ability to detect weakly emitting Be stars, this is still an efficient
method for identifying Be stars in open clusters.  Applied consistently 
to a large sample of open clusters, the variability of the Be phenomenon 
can be neglected and this technique can provide important statistical 
results about these stars, especially regarding their evolutionary status 
and the cause of their rapid rotation.  In an upcoming paper, we will 
apply our photometric method to more than 50 additional clusters and 
discuss the results.

%In an upcoming paper, we will apply our photometric technique to search 
%for Be stars in more than 50 additional clusters.  Although this 
%technique has limitations, ours will be the most extensive survey of 
%Be stars in open clusters that relies on a single, consistent detection 
%technique.  With such a large sample of Be stars in open clusters, we will 
%use our results to examine the evolutionary status of Be stars and the 
%cause of their rapid rotation.

%%%%%%%%%%%%%%%%%%%%%%%%%%%%%%%%%%%%%%%%%%%%%%%%%%%%%%%%%%%%%%%%%%%%%

\acknowledgments

We are grateful to the SMARTS Consortium for our observing time with 
the CTIO 1.5m telescope.  MVM thanks NOAO for travel support to 
observe with the CTIO 0.9m telescope.  MVM is supported by an NSF
Astronomy and Astrophysics Postdoctoral Fellowship under award 
AST$-$0401460.  Financial support also was provided by the NSF through 
grant AST$-$0205297 (DRG).  Institutional support has been provided from 
the GSU College of Arts and Sciences and from the Research Program 
Enhancement fund of the Board of Regents of the University System of 
Georgia, administered through the GSU Office of the Vice President for 
Research.

%%%%%%%%%%%%%%%%%%%%%%%%%%%%%%%%%%%%%%%%%%%%%%%%%%%%%%%%%%%%%%%%%%%%%


%% copy editing.  Individual instruments can be provided in parentheses,
%% after the keyword, but they will not be verified.

Facilities: \facility{CTIO}.

%%%%%%%%%%%%%%%%%%%%%%%%%%%%%%%%%%%%%%%%%%%%%%%%%%%%%%%%%%%%%%%%%%

\clearpage

%%%%%%%%%%%%%%%%%%%%%%%%%%%%%%%%%%%%%%%%%%%%%%%%%%%%%%%%%%%%%%%%%%

%% the file, uncomment the \plotone, \plottwo, and
%% If you need a layout that cannot be achieved with \plotone or
%% \plottwo, you can invoke the graphicx package directly with the
%% \includegraphics command or use \plotfiddle. For more information,
%% please see the tutorial on "Using Electronic Art with AASTeX" in the
%% documentation section at the AASTeX Web site,
%% http://www.journals.uchicago.edu/AAS/AASTeX.

\clearpage

% Figure 1
\notetoeditor{Figures 1, 2, and 3 should be enlarged for visibility.}

% Figure 2

% Figure 3

% Figure 4

\clearpage

% Figure 5

% Figure 6

%%%%%%%%%%%%%%%%%%%%%%%%%%%%%%%%%%%%%%%%%%%%%%%%%%%%%%%%%%%%%%%%%%%

%% Tables should be submitted one per page, so put a \clearpage before

%% Two options are available to the author for producing tables:  the
%% deluxetable environment provided by the AASTeX package or the LaTeX
%% table environment.  Use of deluxetable is preferred.

%% Three table samples follow, two marked up in the deluxetable environment,
%% one marked up as a LaTeX table.

%  example tables & more comments removed, see sample.tex.

\clearpage
\begin{deluxetable}{rrrrcrcrccrl}
\rotate
\tabletypesize{\scriptsize}
\tablecaption{Photometry of NGC 3766 \label{table1}}
\tablehead{
\colhead{Star} &
\colhead{RA (2000)} &
\colhead{Dec (2000)} &
\colhead{$y$~~} &
\colhead{$\delta y$} &
\colhead{$b-y$} &
\colhead{$\delta (b-y)$} &
\colhead{$y-{\rm H}\alpha$} &
\colhead{$\delta (y-{\rm H}\alpha)$} &
\colhead{Code} &
\colhead{WEBDA} &
\colhead{Identifier}
}

\startdata
       1  &   11 35 59.26  &   $-$61 29 14.3  &  12.787  &  0.034  &   0.087  &  0.043  &   0.119  &  0.075  &     O  & \nodata &                              \nodata  \\
       2  &   11 35 48.57  &   $-$61 29 19.2  &   9.708  &  0.011  &   0.047  &  0.016  &   0.174  &  0.031  &     B  & \nodata &                              \nodata  \\
       3  &   11 36 56.50  &   $-$61 29 18.9  &  12.912  &  0.035  &   0.351  &  0.047  &   0.498  &  0.069  &     O  & \nodata &                              \nodata  \\
       4  &   11 36 18.84  &   $-$61 29 32.3  &  12.856  &  0.035  &   0.193  &  0.045  &   0.295  &  0.073  &     O  &   156~~  &                              \nodata  \\
       5  &   11 35 14.04  &   $-$61 29 46.4  &  12.002  &  0.024  &   0.222  &  0.033  &   0.233  &  0.054  &     O  & \nodata &                              \nodata  \\
\hline
       6  &   11 36 43.71  &   $-$61 30 04.1  &  11.866  &  0.025  &   0.844  &  0.039  &   0.987  &  0.044  &     O  & \nodata &                              \nodata  \\
       7  &   11 35 28.04  &   $-$61 30 24.7  &  12.108  &  0.027  &   0.698  &  0.040  &   0.880  &  0.049  &     O  & \nodata &                              \nodata  \\
       8  &   11 35 59.95  &   $-$61 30 26.6  &  11.484  &  0.020  &   0.090  &  0.026  &   0.146  &  0.047  &     B  & \nodata &                              \nodata  \\
       9  &   11 36 23.12  &   $-$61 30 28.9  &  13.078  &  0.041  &   1.199  &  0.066  &   1.360  &  0.062  &     O  & \nodata &                              \nodata  \\
      10  &   11 35 41.55  &   $-$61 30 39.6  &  13.384  &  0.059  &   0.505  &  0.081  &   0.684  &  0.106  &     O  &   303~~  &                              \nodata  \\
\hline
      11  &   11 36 15.81  &   $-$61 30 50.3  &  11.763  &  0.022  &   0.085  &  0.029  &   0.166  &  0.049  &     B  & \nodata &                              \nodata  \\
      12  &   11 35 53.33  &   $-$61 30 59.6  &  13.095  &  0.039  &   0.665  &  0.055  &   0.911  &  0.066  &     O  & \nodata &                              \nodata  \\
      13  &   11 36 08.00  &   $-$61 31 00.6  &  12.751  &  0.033  &   0.409  &  0.045  &   0.576  &  0.064  &     O  & \nodata &                              \nodata  \\
      14  &   11 36 55.10  &   $-$61 31 06.3  &  12.319  &  0.029  &   0.680  &  0.042  &   0.866  &  0.051  &     O  & \nodata &                              \nodata  \\
      15  &   11 36 17.65  &   $-$61 31 27.1  &  11.918  &  0.026  &   0.723  &  0.039  &   0.900  &  0.047  &     O  &   160~~  &                              \nodata  \\
\hline
      16  &   11 36 41.89  &   $-$61 31 37.3  &   9.145  &  0.011  &   0.044  &  0.016  &   0.154  &  0.030  &     B  &   169~~  &                            HD 100969  \\
      17  &   11 35 26.03  &   $-$61 31 39.0  &  13.594  &  0.049  &   1.214  &  0.077  &   1.401  &  0.074  &     O  &   294~~  &                              \nodata  \\
      18  &   11 36 54.78  &   $-$61 31 49.5  &  12.564  &  0.033  &   0.935  &  0.051  &   1.109  &  0.054  &     O  &   173~~  &                              \nodata  \\
      19  &   11 36 06.15  &   $-$61 31 52.9  &  10.210  &  0.017  &   0.679  &  0.027  &   0.873  &  0.033  &     O  &   147~~  &                              \nodata  \\
      20  &   11 36 09.17  &   $-$61 31 55.7  &  11.386  &  0.019  &   0.081  &  0.025  &   0.169  &  0.045  &     B  &   148~~  &                              \nodata  \\
\hline
      21  &   11 36 43.90  &   $-$61 31 59.0  &  13.197  &  0.041  &   0.878  &  0.062  &   1.063  &  0.069  &     O  &   170~~  &                              \nodata  \\
      22  &   11 35 31.46  &   $-$61 32 07.1  &  12.415  &  0.029  &   0.160  &  0.038  &   0.167  &  0.064  &     O  & \nodata &                              \nodata  \\
      23  &   11 36 56.37  &   $-$61 32 08.5  &  12.768  &  0.033  &   0.119  &  0.043  &   0.213  &  0.069  &     B  & \nodata &                              \nodata  \\
      24  &   11 36 07.96  &   $-$61 32 10.5  &  13.311  &  0.041  &   0.837  &  0.062  &   0.983  &  0.071  &     O  &   149~~  &                              \nodata  \\
      25  &   11 35 22.07  &   $-$61 32 10.9  &  10.736  &  0.016  &   0.062  &  0.021  &   0.179  &  0.038  &     B  &   291~~  &                            HD 306793  \\
\hline
      26  &   11 36 00.76  &   $-$61 32 22.7  &  11.752  &  0.022  &   0.084  &  0.028  &   0.173  &  0.050  &     B  &   145~~  &                              \nodata  \\
      27  &   11 35 55.73  &   $-$61 32 32.0  &  11.152  &  0.018  &   0.084  &  0.024  &   0.183  &  0.041  &     B  &   146~~  &                              \nodata  \\
      28  &   11 35 48.24  &   $-$61 32 33.9  &  12.288  &  0.026  &   0.102  &  0.034  &   0.130  &  0.060  &     B  &   134~~  &                              \nodata  \\
      29  &   11 36 12.51  &   $-$61 32 40.2  &  10.281  &  0.014  &   0.050  &  0.020  &   0.147  &  0.035  &     B  & \nodata &                              \nodata  \\
      30  &   11 35 28.63  &   $-$61 32 42.3  &  12.840  &  0.034  &   0.113  &  0.044  &   0.130  &  0.076  &     B  &   292~~  &                              \nodata  \\
\hline
      31  &   11 36 12.35  &   $-$61 32 44.9  &  10.089  &  0.013  &   0.093  &  0.018  &   0.191  &  0.034  &     B  &   151~~  &                 CD $-60^{\circ}3626$  \\
      32  &   11 35 05.96  &   $-$61 33 08.3  &  13.253  &  0.060  &   0.875  &  0.089  &   1.004  &  0.098  &     O  & \nodata &                              \nodata  \\
      33  &   11 36 37.74  &   $-$61 33 10.9  &  10.991  &  0.016  &   0.055  &  0.022  &   0.156  &  0.039  &     B  &   177~~  &                              \nodata  \\
      34  &   11 36 04.39  &   $-$61 33 16.6  &  12.612  &  0.031  &   0.086  &  0.040  &   0.105  &  0.070  &     B  & \nodata &                              \nodata  \\
      35  &   11 35 49.54  &   $-$61 33 18.3  &  12.719  &  0.033  &   0.109  &  0.042  &   0.146  &  0.072  &     B  &   135~~  &                              \nodata  \\
\hline
      36  &   11 36 22.69  &   $-$61 33 33.9  &  12.641  &  0.030  &   0.106  &  0.039  &   0.139  &  0.068  &     B  &    13~~  &                              \nodata  \\
      37  &   11 36 20.73  &   $-$61 33 41.3  &  12.563  &  0.030  &   0.110  &  0.038  &   0.153  &  0.066  &     B  &    12~~  &                              \nodata  \\
      38  &   11 36 11.25  &   $-$61 33 47.4  &  12.260  &  0.027  &   0.120  &  0.035  &   0.100  &  0.061  &     B  &     9~~  &                              \nodata  \\
      39  &   11 36 03.20  &   $-$61 33 54.3  &  13.096  &  0.039  &   0.710  &  0.057  &   0.898  &  0.068  &     O  & \nodata &                              \nodata  \\
      40  &   11 36 51.74  &   $-$61 34 03.7  &  12.018  &  0.024  &   0.081  &  0.032  &   0.172  &  0.055  &     B  &   181~~  &                              \nodata  \\
\hline
      41  &   11 35 36.88  &   $-$61 34 04.9  &  10.619  &  0.015  &   0.050  &  0.021  &   0.151  &  0.037  &     B  &   130~~  &                CPD $-60^{\circ}3088$  \\
      42  &   11 36 38.40  &   $-$61 34 10.5  &  11.409  &  0.019  &   0.054  &  0.025  &   0.147  &  0.045  &     B  &   178~~  &                              \nodata  \\
      43  &   11 36 00.95  &   $-$61 34 17.4  &  11.733  &  0.021  &   0.108  &  0.028  &   0.143  &  0.049  &     B  &   140~~  &                              \nodata  \\
      44  &   11 36 44.23  &   $-$61 34 18.2  &  13.513  &  0.046  &   0.793  &  0.067  &   0.954  &  0.079  &     O  &   180~~  &                              \nodata  \\
      45  &   11 36 08.24  &   $-$61 34 19.3  &  10.685  &  0.015  &   0.049  &  0.021  &   0.140  &  0.037  &     B  &     8~~  &                              \nodata  \\
\hline
      46  &   11 36 40.44  &   $-$61 34 22.7  &  11.464  &  0.019  &   0.098  &  0.025  &   0.161  &  0.044  &     B  &   179~~  &                              \nodata  \\
      47  &   11 36 31.56  &   $-$61 34 25.6  &   8.528  &  0.009  &   0.076  &  0.014  &   0.470  &  0.028  &    Be  &    15~~  &                            HD 306791  \\
      48  &   11 36 00.04  &   $-$61 34 27.0  &  12.172  &  0.026  &   0.084  &  0.034  &   0.146  &  0.058  &     B  &   139~~  &                              \nodata  \\
      49  &   11 35 54.42  &   $-$61 34 30.4  &  10.809  &  0.015  &   0.054  &  0.021  &   0.145  &  0.037  &     B  &   137~~  &                              \nodata  \\
      50  &   11 35 23.43  &   $-$61 34 44.7  &  12.225  &  0.028  &   0.763  &  0.042  &   0.936  &  0.049  &     O  & \nodata &                              \nodata  \\
\hline
      51  &   11 36 31.86  &   $-$61 34 46.8  &   9.936  &  0.011  &   0.003  &  0.016  &   0.145  &  0.032  &     B  &    16~~  &                CPD $-60^{\circ}3158$  \\
      52  &   11 36 04.92  &   $-$61 34 49.1  &  10.074  &  0.012  &   0.058  &  0.018  &   0.150  &  0.032  &     B  &     7~~  &                              \nodata  \\
      53  &   11 36 20.68  &   $-$61 34 59.4  &  12.359  &  0.027  &   0.133  &  0.035  &   0.179  &  0.059  &     B  &    19~~  &                              \nodata  \\
      54  &   11 35 35.86  &   $-$61 35 00.7  &  10.361  &  0.014  &   0.058  &  0.020  &   0.161  &  0.035  &     B  &   125~~  &                CPD $-60^{\circ}3083$  \\
      55  &   11 36 15.87  &   $-$61 35 01.6  &  10.402  &  0.014  &   0.057  &  0.019  &   0.153  &  0.035  &     B  & \nodata &                              \nodata  \\
\hline
      56  &   11 35 44.89  &   $-$61 35 05.0  &  12.595  &  0.031  &   0.101  &  0.040  &   0.167  &  0.068  &     B  &   127~~  &                              \nodata  \\
      57  &   11 36 12.03  &   $-$61 35 10.6  &  11.289  &  0.019  &   0.068  &  0.025  &   0.167  &  0.044  &     B  &    24~~  &                CPD $-60^{\circ}3134$  \\
      58  &   11 35 14.99  &   $-$61 35 10.2  &  12.235  &  0.027  &   0.100  &  0.035  &   0.098  &  0.061  &     B  &   281~~  &                              \nodata  \\
      59  &   11 35 47.72  &   $-$61 35 11.1  &  11.955  &  0.025  &   0.817  &  0.039  &   0.988  &  0.045  &     O  &   129~~  &                              \nodata  \\
      60  &   11 36 29.57  &   $-$61 35 14.0  &  12.811  &  0.033  &   0.117  &  0.043  &   0.171  &  0.073  &     B  &    17~~  &                              \nodata  \\
\hline
      61  &   11 36 21.38  &   $-$61 35 15.0  &   9.581  &  0.012  &   0.072  &  0.018  &   0.218  &  0.031  &   Be?  &    20~~  &                              \nodata  \\
      62  &   11 35 46.47  &   $-$61 35 17.9  &  13.246  &  0.041  &   0.474  &  0.056  &   0.647  &  0.077  &     O  &   128~~  &                              \nodata  \\
      63  &   11 35 43.53  &   $-$61 35 18.6  &  13.644  &  0.048  &   0.519  &  0.066  &   0.955  &  0.082  &     O  &   126~~  &                              \nodata  \\
      64  &   11 36 25.51  &   $-$61 35 19.0  &  12.587  &  0.029  &   0.119  &  0.038  &   0.182  &  0.067  &     B  &    18~~  &                              \nodata  \\
      65  &   11 35 21.21  &   $-$61 35 20.4  &  11.777  &  0.022  &   0.099  &  0.029  &   0.166  &  0.051  &     B  & \nodata &                              \nodata  \\
\hline
      66  &   11 36 20.77  &   $-$61 35 21.4  &  11.731  &  0.022  &   0.223  &  0.030  &   0.265  &  0.049  &     O  &  1076~~  &                              \nodata  \\
      67  &   11 36 08.97  &   $-$61 35 21.7  &  11.173  &  0.018  &   0.066  &  0.024  &   0.171  &  0.042  &     B  &    25~~  &                              \nodata  \\
      68  &   11 36 15.19  &   $-$61 35 22.4  &  12.083  &  0.025  &   0.081  &  0.033  &   0.103  &  0.058  &     B  & \nodata &                              \nodata  \\
      69  &   11 36 04.55  &   $-$61 35 23.0  &   8.174  &  0.013  &   0.073  &  0.021  &   0.167  &  0.033  &     B  & \nodata &                              \nodata  \\
      70  &   11 36 59.16  &   $-$61 35 31.4  &  12.889  &  0.050  &   0.609  &  0.071  &   0.935  &  0.084  &     O  & \nodata &                              \nodata  \\
\hline
      71  &   11 36 26.36  &   $-$61 35 36.8  &  11.533  &  0.019  &   0.071  &  0.025  &   0.142  &  0.045  &     B  & \nodata &                              \nodata  \\
      72  &   11 36 01.98  &   $-$61 35 38.1  &  12.638  &  0.031  &   0.110  &  0.040  &   0.162  &  0.068  &     B  &     4~~  &                              \nodata  \\
      73  &   11 36 09.56  &   $-$61 35 38.1  &   9.228  &  0.011  &   0.062  &  0.016  &   0.218  &  0.030  &     B  &    26~~  &                              \nodata  \\
      74  &   11 36 21.24  &   $-$61 35 38.4  &  10.661  &  0.015  &   0.057  &  0.021  &   0.162  &  0.037  &     B  &  1031~~  &                              \nodata  \\
      75  &   11 35 25.69  &   $-$61 35 38.7  &  12.467  &  0.029  &   0.109  &  0.038  &   0.139  &  0.065  &     B  &   121~~  &                              \nodata  \\
\hline
      76  &   11 36 22.05  &   $-$61 35 38.8  &   9.115  &  0.011  &   0.071  &  0.016  &   0.171  &  0.030  &     B  &    21~~  &                              \nodata  \\
      77  &   11 36 40.81  &   $-$61 35 41.1  &  11.357  &  0.019  &   0.096  &  0.025  &   0.177  &  0.043  &     B  &   195~~  &                              \nodata  \\
      78  &   11 36 18.29  &   $-$61 35 42.3  &   9.983  &  0.013  &   0.055  &  0.018  &   0.152  &  0.034  &     B  &    22~~  &                CPD $-60^{\circ}3137$  \\
      79  &   11 35 32.66  &   $-$61 35 42.3  &  11.131  &  0.016  &   0.064  &  0.022  &   0.165  &  0.040  &     B  &   118~~  &                              \nodata  \\
      80  &   11 36 08.59  &   $-$61 35 42.5  &  12.720  &  0.033  &   0.162  &  0.043  &   0.212  &  0.071  &     O  &  1147~~  &                              \nodata  \\
\hline
      81  &   11 35 50.71  &   $-$61 35 46.8  &  12.772  &  0.033  &   0.120  &  0.043  &   0.161  &  0.073  &     B  & \nodata &                              \nodata  \\
      82  &   11 36 10.08  &   $-$61 35 47.2  &  12.868  &  0.033  &   0.096  &  0.043  &   0.133  &  0.075  &     B  &  1164~~  &                              \nodata  \\
      83  &   11 36 11.86  &   $-$61 35 50.2  &   8.337  &  0.010  &   0.072  &  0.016  &   0.246  &  0.029  &   Be?  &    27~~  &                CPD $-60^{\circ}3128$  \\
      84  &   11 36 03.57  &   $-$61 35 50.6  &  11.485  &  0.020  &   0.054  &  0.026  &   0.159  &  0.047  &     B  &     6~~  &                              \nodata  \\
      85  &   11 36 27.33  &   $-$61 35 52.7  &  12.774  &  0.033  &   0.100  &  0.042  &   0.089  &  0.074  &     B  &    40~~  &                              \nodata  \\
\hline
      86  &   11 36 10.44  &   $-$61 35 59.0  &  11.178  &  0.018  &   0.076  &  0.024  &   0.150  &  0.043  &     B  &    28~~  &                              \nodata  \\
      87  &   11 35 44.07  &   $-$61 36 04.4  &  11.635  &  0.021  &   0.082  &  0.028  &   0.169  &  0.049  &     B  &   114~~  &                              \nodata  \\
      88  &   11 35 48.87  &   $-$61 36 06.7  &  10.347  &  0.014  &   0.386  &  0.021  &   0.534  &  0.033  &     O  & \nodata &                              \nodata  \\
      89  &   11 36 15.78  &   $-$61 36 08.1  &  11.086  &  0.017  &   0.073  &  0.023  &   0.144  &  0.042  &     B  & \nodata &                              \nodata  \\
      90  &   11 36 26.84  &   $-$61 36 10.6  &  11.715  &  0.022  &   0.108  &  0.029  &   0.105  &  0.050  &     B  & \nodata &                              \nodata  \\
\hline
      91  &   11 36 06.96  &   $-$61 36 12.6  &  12.872  &  0.033  &   0.105  &  0.043  &   0.156  &  0.073  &     B  & \nodata &                              \nodata  \\
      92  &   11 35 55.45  &   $-$61 36 13.7  &   8.615  &  0.011  &   0.064  &  0.016  &   0.219  &  0.029  &   Be?  &     1~~  &                            HD 100856  \\
      93  &   11 36 32.03  &   $-$61 36 17.2  &  12.683  &  0.033  &   0.481  &  0.045  &   0.529  &  0.064  &     O  &    44~~  &                              \nodata  \\
      94  &   11 36 49.34  &   $-$61 36 20.9  &  10.791  &  0.016  &   0.104  &  0.022  &   0.272  &  0.038  &   Be?  &   194~~  &                              \nodata  \\
      95  &   11 35 44.34  &   $-$61 36 23.3  &  12.144  &  0.024  &   0.098  &  0.032  &   0.152  &  0.056  &     B  &   113~~  &                              \nodata  \\
\hline
      96  &   11 36 32.10  &   $-$61 36 25.9  &  11.610  &  0.021  &   0.106  &  0.028  &   0.165  &  0.048  &     B  &    45~~  &                              \nodata  \\
      97  &   11 36 25.59  &   $-$61 36 26.1  &  11.892  &  0.022  &   0.102  &  0.029  &   0.141  &  0.052  &     B  &    42~~  &                              \nodata  \\
      98  &   11 36 21.85  &   $-$61 36 29.8  &  10.587  &  0.015  &   0.066  &  0.021  &   0.174  &  0.036  &     B  &    36~~  &                             V848 Cen  \\
      99  &   11 36 09.07  &   $-$61 36 33.3  &  12.098  &  0.025  &   0.080  &  0.033  &   0.167  &  0.057  &     B  &    32~~  &                              \nodata  \\
     100  &   11 36 32.16  &   $-$61 36 38.0  &  12.765  &  0.033  &   0.042  &  0.042  &   0.114  &  0.075  &     B  & \nodata &                              \nodata  \\
\hline
     101  &   11 36 10.86  &   $-$61 36 38.8  &  12.287  &  0.027  &   0.083  &  0.035  &   0.156  &  0.061  &     B  &    34~~  &                              \nodata  \\
     102  &   11 36 23.21  &   $-$61 36 40.4  &  11.566  &  0.019  &   0.089  &  0.025  &   0.174  &  0.045  &     B  &    38~~  &                              \nodata  \\
     103  &   11 35 35.38  &   $-$61 36 41.0  &  11.373  &  0.018  &   0.071  &  0.024  &   0.159  &  0.044  &     B  &   111~~  &                              \nodata  \\
     104  &   11 35 53.63  &   $-$61 36 42.6  &  12.785  &  0.033  &   0.155  &  0.043  &   0.166  &  0.073  &     O  &   109~~  &                              \nodata  \\
     105  &   11 36 09.43  &   $-$61 36 43.0  &  12.324  &  0.027  &   0.094  &  0.035  &   0.116  &  0.060  &     B  &    33~~  &                              \nodata  \\
\hline
     106  &   11 36 40.34  &   $-$61 36 45.4  &  12.861  &  0.037  &   1.073  &  0.059  &   1.248  &  0.059  &     O  & \nodata &                              \nodata  \\
     107  &   11 36 19.02  &   $-$61 36 45.7  &  12.203  &  0.027  &   0.105  &  0.035  &   0.188  &  0.059  &     B  & \nodata &                              \nodata  \\
     108  &   11 36 30.11  &   $-$61 36 47.6  &  12.269  &  0.027  &   0.104  &  0.035  &   0.126  &  0.061  &     B  &    43~~  &                              \nodata  \\
     109  &   11 36 59.44  &   $-$61 36 49.8  &  10.920  &  0.023  &   0.066  &  0.032  &   0.159  &  0.051  &     B  &   193~~  &                              \nodata  \\
     110  &   11 35 41.32  &   $-$61 36 56.2  &  10.520  &  0.013  &   0.048  &  0.018  &   0.140  &  0.035  &     B  &   107~~  &                CPD $-60^{\circ}3091$  \\
\hline
     111  &   11 36 21.01  &   $-$61 36 58.3  &  10.236  &  0.013  &   0.078  &  0.018  &   0.166  &  0.034  &     B  &    52~~  &                              \nodata  \\
     112  &   11 36 18.85  &   $-$61 37 03.8  &  11.149  &  0.018  &   0.081  &  0.024  &   0.183  &  0.042  &     B  & \nodata &                              \nodata  \\
     113  &   11 36 11.31  &   $-$61 37 07.2  &  12.763  &  0.033  &   0.111  &  0.043  &   0.177  &  0.073  &     B  &    61~~  &                              \nodata  \\
     114  &   11 36 16.30  &   $-$61 37 07.4  &  12.339  &  0.028  &   0.105  &  0.036  &   0.176  &  0.062  &     B  & \nodata &                              \nodata  \\
     115  &   11 36 17.74  &   $-$61 37 08.6  &  11.554  &  0.021  &   0.106  &  0.028  &   0.165  &  0.048  &     B  & \nodata &                              \nodata  \\
\hline
     116  &   11 36 18.75  &   $-$61 37 10.2  &  12.174  &  0.026  &   0.105  &  0.034  &   0.201  &  0.057  &     B  & \nodata &                              \nodata  \\
     117  &   11 35 38.71  &   $-$61 37 13.7  &  13.235  &  0.041  &   0.666  &  0.059  &   0.881  &  0.072  &     O  &   105~~  &                              \nodata  \\
     118  &   11 36 33.08  &   $-$61 37 14.4  &  12.019  &  0.023  &   0.088  &  0.030  &   0.188  &  0.054  &     B  & \nodata &                              \nodata  \\
     119  &   11 36 01.10  &   $-$61 37 18.4  &  10.099  &  0.012  &   0.065  &  0.018  &   0.164  &  0.033  &     B  &    81~~  &                              \nodata  \\
     120  &   11 36 25.96  &   $-$61 37 19.0  &  12.699  &  0.032  &   0.131  &  0.042  &   0.144  &  0.072  &     B  &    51~~  &                              \nodata  \\
\hline
     121  &   11 35 47.91  &   $-$61 37 19.6  &  12.242  &  0.027  &   0.086  &  0.034  &   0.173  &  0.058  &     B  &   108~~  &                              \nodata  \\
     122  &   11 36 10.74  &   $-$61 37 21.0  &  12.231  &  0.025  &   0.136  &  0.033  &   0.088  &  0.059  &     B  &    62~~  &                              \nodata  \\
     123  &   11 36 55.73  &   $-$61 37 21.6  &  11.732  &  0.022  &   0.317  &  0.030  &   0.472  &  0.046  &     O  & \nodata &                              \nodata  \\
     124  &   11 36 02.65  &   $-$61 37 24.6  &  12.465  &  0.029  &   0.126  &  0.038  &   0.024  &  0.068  &     B  &  1200~~  &                              \nodata  \\
     125  &   11 36 14.40  &   $-$61 37 24.7  &  11.962  &  0.023  &   0.100  &  0.030  &   0.163  &  0.052  &     B  &    66~~  &                              \nodata  \\
\hline
     126  &   11 36 02.42  &   $-$61 37 26.4  &  12.470  &  0.029  &   0.068  &  0.038  &   0.218  &  0.064  &   Be?  & \nodata &                              \nodata  \\
     127  &   11 36 21.94  &   $-$61 37 28.4  &  10.847  &  0.016  &   0.084  &  0.022  &   0.405  &  0.038  &    Be  &    53~~  &                              \nodata  \\
     128  &   11 35 35.73  &   $-$61 37 30.1  &  12.238  &  0.027  &   0.071  &  0.035  &   0.158  &  0.059  &     B  &   103~~  &                              \nodata  \\
     129  &   11 35 41.15  &   $-$61 37 32.5  &  12.329  &  0.027  &   0.103  &  0.035  &   0.179  &  0.059  &     B  & \nodata &                              \nodata  \\
     130  &   11 36 14.05  &   $-$61 37 35.6  &   9.927  &  0.013  &   0.070  &  0.018  &   0.266  &  0.033  &   Be?  &    67~~  &                              \nodata  \\
\hline
     131  &   11 36 21.64  &   $-$61 37 38.0  &  11.376  &  0.019  &   0.090  &  0.025  &   0.171  &  0.045  &     B  &    54~~  &                              \nodata  \\
     132  &   11 36 46.91  &   $-$61 37 39.0  &  13.398  &  0.044  &   0.684  &  0.063  &   0.910  &  0.076  &     O  &  1238~~  &                              \nodata  \\
     133  &   11 36 10.16  &   $-$61 37 39.8  &   9.241  &  0.010  &   0.068  &  0.016  &   0.178  &  0.030  &     B  &    63~~  &                             V846 Cen  \\
     134  &   11 36 47.27  &   $-$61 37 42.2  &  12.590  &  0.031  &   0.106  &  0.040  &   0.214  &  0.068  &     B  & \nodata &                              \nodata  \\
     135  &   11 36 02.18  &   $-$61 37 43.6  &  12.872  &  0.035  &   0.128  &  0.045  &   0.150  &  0.076  &     B  &    84~~  &                              \nodata  \\
\hline
     136  &   11 36 08.41  &   $-$61 37 47.5  &  11.870  &  0.022  &   0.140  &  0.029  &   0.174  &  0.051  &     B  &    64~~  &                              \nodata  \\
     137  &   11 36 17.42  &   $-$61 37 48.4  &  12.120  &  0.024  &   0.138  &  0.032  &   0.205  &  0.055  &     B  & \nodata &                              \nodata  \\
     138  &   11 36 05.84  &   $-$61 37 51.4  &  11.442  &  0.018  &   0.088  &  0.024  &   0.172  &  0.044  &     B  & \nodata &                              \nodata  \\
     139  &   11 36 41.89  &   $-$61 37 53.7  &  11.050  &  0.017  &   0.084  &  0.023  &   0.227  &  0.040  &   Be?  &   204~~  &                              \nodata  \\
     140  &   11 36 23.89  &   $-$61 38 02.9  &  12.848  &  0.033  &   0.157  &  0.043  &   0.141  &  0.074  &     O  &    69~~  &                              \nodata  \\
\hline
     141  &   11 35 45.12  &   $-$61 38 03.4  &  12.822  &  0.033  &   0.145  &  0.043  &   0.156  &  0.074  &     B  &    99~~  &                              \nodata  \\
     142  &   11 35 08.53  &   $-$61 38 06.9  &  11.762  &  0.031  &   0.076  &  0.041  &   0.167  &  0.066  &     B  & \nodata &                              \nodata  \\
     143  &   11 35 52.25  &   $-$61 38 07.5  &   9.634  &  0.012  &   0.056  &  0.018  &   0.168  &  0.031  &     B  &    97~~  &                CPD $-60^{\circ}3098$  \\
     144  &   11 36 51.60  &   $-$61 38 08.2  &  11.284  &  0.026  &   1.199  &  0.044  &   1.300  &  0.040  &     O  &   202~~  &                              \nodata  \\
     145  &   11 36 38.82  &   $-$61 38 11.7  &  12.215  &  0.027  &   0.125  &  0.035  &   0.161  &  0.059  &     B  &   205~~  &                              \nodata  \\
\hline
     146  &   11 36 08.73  &   $-$61 38 12.8  &  10.669  &  0.014  &   0.152  &  0.021  &   0.179  &  0.036  &     O  & \nodata &                              \nodata  \\
     147  &   11 35 52.37  &   $-$61 38 15.2  &  11.839  &  0.023  &   0.094  &  0.030  &   0.193  &  0.051  &     B  &  1086~~  &                              \nodata  \\
     148  &   11 35 57.02  &   $-$61 38 15.6  &  11.213  &  0.018  &   0.078  &  0.025  &   0.189  &  0.043  &     B  &    87~~  &                              \nodata  \\
     149  &   11 36 59.58  &   $-$61 38 22.3  &  12.021  &  0.033  &   0.127  &  0.043  &   0.144  &  0.074  &     B  &   208~~  &                              \nodata  \\
     150  &   11 35 31.33  &   $-$61 38 24.5  &  12.485  &  0.028  &   0.119  &  0.037  &   0.155  &  0.063  &     B  & \nodata &                              \nodata  \\
\hline
     151  &   11 35 46.89  &   $-$61 38 25.3  &  12.060  &  0.025  &   0.081  &  0.033  &   0.162  &  0.056  &     B  &    98~~  &                              \nodata  \\
     152  &   11 36 11.07  &   $-$61 38 28.5  &  11.243  &  0.017  &   0.083  &  0.023  &   0.145  &  0.042  &     B  & \nodata &                              \nodata  \\
     153  &   11 36 13.20  &   $-$61 38 36.4  &  12.918  &  0.036  &   0.672  &  0.052  &   0.854  &  0.064  &     O  &    77~~  &                              \nodata  \\
     154  &   11 36 08.02  &   $-$61 38 38.2  &   9.985  &  0.013  &   0.071  &  0.018  &   0.663  &  0.032  &    Be  &    88~~  &                             V845 Cen  \\
     155  &   11 36 46.00  &   $-$61 38 37.7  &  12.362  &  0.028  &   0.117  &  0.036  &   0.182  &  0.062  &     B  & \nodata &                              \nodata  \\
\hline
     156  &   11 36 36.98  &   $-$61 38 42.8  &  12.792  &  0.033  &   0.175  &  0.043  &   0.143  &  0.073  &     O  &   206~~  &                              \nodata  \\
     157  &   11 36 05.44  &   $-$61 38 48.8  &  12.850  &  0.034  &   0.133  &  0.044  &   0.183  &  0.075  &     B  &    89~~  &                              \nodata  \\
     158  &   11 36 04.27  &   $-$61 38 53.8  &  11.886  &  0.023  &   0.085  &  0.030  &   0.165  &  0.052  &     B  &    90~~  &                              \nodata  \\
     159  &   11 35 59.53  &   $-$61 38 58.6  &  11.776  &  0.021  &   0.096  &  0.028  &   0.155  &  0.049  &     B  &    93~~  &                              \nodata  \\
     160  &   11 36 48.96  &   $-$61 38 58.5  &  12.980  &  0.051  &   0.133  &  0.065  &   0.144  &  0.107  &     B  & \nodata &                              \nodata  \\
\hline
     161  &   11 36 21.10  &   $-$61 39 03.3  &   9.084  &  0.011  &   0.056  &  0.016  &   0.167  &  0.030  &     B  &    70~~  &                 CD $-60^{\circ}3629$  \\
     162  &   11 36 34.47  &   $-$61 39 13.2  &  12.581  &  0.031  &   0.119  &  0.040  &   0.120  &  0.069  &     B  & \nodata &                              \nodata  \\
     163  &   11 36 33.56  &   $-$61 39 14.1  &  13.528  &  0.047  &   0.773  &  0.068  &   0.997  &  0.079  &     O  & \nodata &                              \nodata  \\
     164  &   11 36 03.41  &   $-$61 39 15.3  &  11.844  &  0.023  &   0.087  &  0.030  &   0.157  &  0.052  &     B  &    91~~  &                              \nodata  \\
     165  &   11 36 00.22  &   $-$61 39 16.8  &  12.302  &  0.026  &   0.115  &  0.034  &   0.129  &  0.059  &     B  & \nodata &                              \nodata  \\
\hline
     166  &   11 36 37.72  &   $-$61 39 23.4  &  12.470  &  0.029  &   0.107  &  0.038  &   0.137  &  0.064  &     B  &   214~~  &                              \nodata  \\
     167  &   11 35 16.16  &   $-$61 39 27.1  &  12.046  &  0.024  &   0.422  &  0.034  &   0.590  &  0.049  &     O  &   270~~  &                              \nodata  \\
     168  &   11 36 28.07  &   $-$61 39 30.8  &  12.829  &  0.033  &   0.116  &  0.043  &   0.150  &  0.074  &     B  &    71~~  &                              \nodata  \\
     169  &   11 35 39.12  &   $-$61 39 37.0  &  12.012  &  0.024  &   0.159  &  0.032  &   0.172  &  0.055  &     O  &   100~~  &                              \nodata  \\
     170  &   11 35 52.00  &   $-$61 39 39.6  &  10.676  &  0.015  &   0.047  &  0.021  &   0.150  &  0.037  &     B  &    94~~  &                            HD 306795  \\
\hline
     171  &   11 36 54.96  &   $-$61 39 39.2  &  11.567  &  0.019  &   0.081  &  0.025  &   0.169  &  0.046  &     B  &   211~~  &                              \nodata  \\
     172  &   11 36 11.09  &   $-$61 39 42.0  &  12.158  &  0.025  &   0.140  &  0.033  &   0.088  &  0.058  &     B  & \nodata &                              \nodata  \\
     173  &   11 36 29.91  &   $-$61 39 47.1  &  12.406  &  0.029  &   0.124  &  0.038  &   0.214  &  0.063  &     B  &   233~~  &                              \nodata  \\
     174  &   11 36 34.27  &   $-$61 39 49.8  &  12.600  &  0.031  &   0.128  &  0.040  &   0.169  &  0.068  &     B  &   217~~  &                              \nodata  \\
     175  &   11 36 36.02  &   $-$61 39 57.9  &  11.848  &  0.023  &   0.093  &  0.030  &   0.178  &  0.052  &     B  &   218~~  &                              \nodata  \\
\hline
     176  &   11 36 50.47  &   $-$61 40 00.8  &   9.133  &  0.011  &   0.053  &  0.016  &   0.144  &  0.030  &     B  &   212~~  &                            HD 100989  \\
     177  &   11 36 50.33  &   $-$61 40 07.6  &  11.924  &  0.024  &   0.377  &  0.033  &   0.506  &  0.049  &     O  &  1088~~  &                              \nodata  \\
     178  &   11 36 48.63  &   $-$61 40 10.0  &  11.883  &  0.023  &   0.080  &  0.030  &   0.143  &  0.053  &     B  &   213~~  &                              \nodata  \\
     179  &   11 36 24.40  &   $-$61 40 23.9  &  10.468  &  0.014  &   0.055  &  0.020  &   0.161  &  0.035  &     B  &   234~~  &                              \nodata  \\
     180  &   11 36 16.50  &   $-$61 40 34.6  &  13.377  &  0.043  &   0.397  &  0.058  &   0.674  &  0.080  &     O  &    74~~  &                              \nodata  \\
\hline
     181  &   11 36 02.94  &   $-$61 40 36.7  &  13.215  &  0.040  &   0.530  &  0.056  &   0.741  &  0.074  &     O  & \nodata &                              \nodata  \\
     182  &   11 35 42.12  &   $-$61 40 39.2  &  11.424  &  0.025  &   0.999  &  0.040  &   1.182  &  0.041  &     O  &   258~~  &                              \nodata  \\
     183  &   11 36 05.73  &   $-$61 40 40.9  &  10.302  &  0.015  &   0.364  &  0.022  &   0.519  &  0.034  &     O  &    95~~  &                              \nodata  \\
     184  &   11 36 18.18  &   $-$61 40 42.0  &  10.837  &  0.015  &   0.080  &  0.021  &   0.147  &  0.037  &     B  & \nodata &                              \nodata  \\
     185  &   11 35 42.89  &   $-$61 40 53.4  &  10.965  &  0.017  &   0.068  &  0.023  &   0.160  &  0.041  &     B  &   257~~  &                            HD 306796  \\
\hline
     186  &   11 36 41.44  &   $-$61 40 58.0  &  12.836  &  0.034  &   0.107  &  0.044  &   0.120  &  0.076  &     B  &   221~~  &                              \nodata  \\
     187  &   11 35 40.30  &   $-$61 41 02.6  &  13.276  &  0.043  &   0.901  &  0.064  &   1.116  &  0.069  &     O  &   259~~  &                              \nodata  \\
     188  &   11 36 37.73  &   $-$61 41 02.5  &  12.556  &  0.031  &   0.230  &  0.040  &   0.109  &  0.068  &     O  &   220~~  &                              \nodata  \\
     189  &   11 36 37.58  &   $-$61 41 04.1  &  12.544  &  0.030  &   0.294  &  0.040  &   0.053  &  0.069  &     O  &  1170~~  &                              \nodata  \\
     190  &   11 35 37.15  &   $-$61 41 09.7  &  11.784  &  0.021  &   0.088  &  0.028  &   0.185  &  0.049  &     B  &   260~~  &                              \nodata  \\
\hline
     191  &   11 36 18.32  &   $-$61 41 12.9  &  13.323  &  0.043  &   1.169  &  0.071  &   1.324  &  0.067  &     O  & \nodata &                              \nodata  \\
     192  &   11 36 12.13  &   $-$61 41 18.7  &  11.526  &  0.019  &   0.120  &  0.025  &   0.146  &  0.046  &     B  & \nodata &                              \nodata  \\
     193  &   11 35 47.47  &   $-$61 41 24.5  &  12.998  &  0.038  &   0.857  &  0.057  &   1.064  &  0.062  &     O  &  1172~~  &                              \nodata  \\
     194  &   11 36 29.59  &   $-$61 41 35.2  &  12.926  &  0.050  &   0.107  &  0.065  &   0.163  &  0.104  &     B  &   231~~  &                              \nodata  \\
     195  &   11 36 21.10  &   $-$61 41 40.6  &  12.684  &  0.031  &   0.164  &  0.041  &   0.219  &  0.068  &     O  & \nodata &                              \nodata  \\
\hline
     196  &   11 36 09.37  &   $-$61 41 41.5  &   9.444  &  0.011  &   0.059  &  0.016  &   0.163  &  0.031  &     B  &   239~~  &                            HD 306798  \\
     197  &   11 35 50.86  &   $-$61 41 56.4  &  12.120  &  0.025  &   0.093  &  0.033  &   0.153  &  0.057  &     B  &   253~~  &                              \nodata  \\
     198  &   11 35 15.15  &   $-$61 41 59.5  &  10.428  &  0.014  &   0.117  &  0.020  &   0.911  &  0.032  &    Be  &   264~~  &                            HD 306657  \\
     199  &   11 35 56.96  &   $-$61 42 04.8  &  13.821  &  0.075  &   0.921  &  0.113  &   0.983  &  0.120  &     O  &   250~~  &                              \nodata  \\
     200  &   11 36 05.48  &   $-$61 42 06.0  &   9.700  &  0.011  &   0.065  &  0.016  &   0.369  &  0.030  &     B  &   240~~  &                            HD 306797  \\
\hline
     201  &   11 36 42.24  &   $-$61 42 13.4  &  11.160  &  0.021  &   0.755  &  0.033  &   0.933  &  0.038  &     O  & \nodata &                              \nodata  \\
     202  &   11 36 58.38  &   $-$61 42 18.8  &  13.493  &  0.066  &   1.072  &  0.104  &   1.679  &  0.091  &     O  &  1390~~  &                              \nodata  \\
     203  &   11 36 58.39  &   $-$61 42 18.9  &  13.498  &  0.067  &   1.132  &  0.106  &   1.684  &  0.092  &     O  &   225~~  &                              \nodata  \\
     204  &   11 35 15.97  &   $-$61 42 20.9  &  12.240  &  0.028  &   0.779  &  0.043  &   0.960  &  0.050  &     O  &   263~~  &                              \nodata  \\
     205  &   11 35 57.08  &   $-$61 42 22.1  &  13.599  &  0.068  &   0.822  &  0.100  &   0.968  &  0.110  &     O  &   249~~  &                              \nodata  \\
\hline
     206  &   11 36 41.64  &   $-$61 42 22.2  &  10.800  &  0.015  &   0.151  &  0.021  &   0.227  &  0.036  &     O  &   229~~  &                              \nodata  \\
     207  &   11 36 37.71  &   $-$61 42 26.5  &  12.016  &  0.027  &   0.915  &  0.043  &   1.102  &  0.046  &     O  & \nodata &                              \nodata  \\
     208  &   11 36 47.25  &   $-$61 42 31.5  &  13.433  &  0.064  &   0.833  &  0.094  &   0.973  &  0.106  &     O  & \nodata &                              \nodata  \\
     209  &   11 35 43.54  &   $-$61 42 47.2  &  13.560  &  0.047  &   0.716  &  0.067  &   0.977  &  0.080  &     O  & \nodata &                              \nodata  \\
\enddata

\tablecomments{
%The complete version of this table is in the electronic 
%edition of the Journal.  The printed edition contains only a sample.  
For each star observed in NGC 3766, we give the right
ascension (RA) and declination (Dec) for the epoch 2000.  We also provide
the $y$ magnitude, the $b-y$ and $y-\rm H\alpha$ colors, and the error
for each.  The code is used to label definite Be stars (Be), possible Be
stars (Be?), B-type stars (B), and other stars in the field (O).  Finally, 
the WEBDA number and other identifiers are provided where available.}

\notetoeditor{Only the first few lines of Table 1 should be included in 
the printed version; the full table should be published in electronic 
format only.  Also, we can submit an ASCII version for the electronic 
edition if necessary.}

\end{deluxetable}

\end{document}